\documentstyle[12pt,epsf]{article}\textheight 230mm\textwidth 160mm
\newcommand{\be}{\begin{equation}}
\newcommand{\ee}{\end{equation}}
\newcommand{\bea}{\begin{eqnarray}}
\newcommand{\eea}{\end{eqnarray}}
\newcommand{\bi}{\bibitem}
\newcommand{\nn}{\nonumber}
\begin{document}
\pagestyle{empty}

\hspace*{11cm} \vspace{-2mm} MPI-PhT/95-132\\
\hspace*{11.7cm} \vspace{-2mm} BUTP-95-37\\
\hspace*{11.7cm}November 1995

\begin{center}{\bf\Large TOP QUARK MASS PREDICTIONS\\
     FROM\\
     GAUGE-YUKAWA UNIFICATION $\ ^{\dag}$}
\end{center}

\begin{center}
     {\it Dedicated to the memory of our very 
good friend Roger Decker}
\end{center}

\begin{center}{\sc Jisuke Kubo}$\ ^{(1),*}$, 
{\sc Myriam
Mondrag{\' o}n}$\ ^{(2),**}$, \vspace{-1mm}
 and 
{\sc George Zoupanos}$\ ^{(3),***}$  
\end{center}

\begin{center}
{\em $\ ^{(1)}$ 
 College of Liberal Arts, Kanazawa \vspace{-2mm} University,
Kanazawa 920-11, Japan } \\
{\em $\ ^{(2)}$ Institut f{\" u}r  \vspace{-2mm} Theoretische Physik,
Philosophenweg 16\\
D-69120 Heidelberg, Germany} \vspace{-2mm}\\
{\em $\ ^{(3)}$ Max-Planck-Institut f\"ur Physik,
 Werner-Heisenberg-Institut \vspace{-2mm}\\
D-80805 Munich, Germany} \vspace{-2mm} 
\\ {\em and} \vspace{-2mm} \\
{\em Institut f{\" u}r Theoretische Physik,
Universit{\" a}t Bern \vspace{-2mm}\\
CH-3012 Bern, Switzerland}
\end{center}

\begin{center}
{\sc\large Abstract}
\end{center}

\noindent
Gauge-Yukawa Unification (GYU) \vspace{-2mm} is a renormalization
group invariant functional relation among 
gauge and Yukawa couplings which holds \vspace{-2mm} beyond the
unification point in Grand Unified Theories (GUTs).
Here, \vspace{-2mm} GYU is obtained by requiring finiteness and
reduction of couplings to all orders in perturbation theory.  We
examine \vspace{-2mm} the consequences of GYU in various
supersymmetric GUTs and \vspace{-2mm} in particular the predictions
for the top quark mass.  These predictions are such that they
distinguish \vspace{-2mm} already GYU from ordinary GUTs.  
Moreover, when more accurate \vspace{-2mm} measurements of the top
quark mass are available, it is expected that it will be possible to
discriminate among the various GYUs.

\vspace*{0.5cm}
\footnoterule
\vspace*{2mm}
\noindent
$\ ^{\dag}$ Presented by \vspace{-3mm} G. Zoupanos at
the '95 Summer School on HEP and Cosmology 
(ICTP, Trieste). \vspace{-3mm} \\
$^{*}$Partially supported  by the Grants-in-Aid
for \vspace{-3mm} Scientific Research  from the Ministry of
Education, Science 
and Culture \vspace{-3mm}  (No. 40211213).\\
\noindent
$^{**}$Address  after Jan. 1st, '96:
Instituto de F{\' \i}sica,  \vspace{-3mm} UNAM,
Apdo. Postal 20-364,
M{\' e}xico 01000 D.F., \vspace{-3mm} M{\' e}xico.\\
\noindent
$ ^{***}$On leave \vspace{-3mm} of absence from 
Physics Department, National
Technical University,  GR-157 80 Zografou, Athens, Greece.  
Partially supported \vspace{-3mm} by C.E.C. projects, SC1-CT91-0729;
CHRX-CT93-0319.

\newpage
\pagestyle{plain}
\section{Introduction}
The apparent success of the standard model (SM) in describing the
elementary particles and their interactions is spoiled by 
the presence of the 
plethora of free parameters.  Therefore, it provides a challenge to 
theorists to attempt to understand at least some of these parameters. 

The traditional way to reduce the independent parameters of a theory
is the introduction of a symmetry.  Grand Unified Theories (GUTs)
\cite {pati1,gut1,GUTS} are
representative examples of such attempts, and some of them are
certainly successful to some extent.  For instance, the minimal
$SU(5)$ reduces by one the gauge couplings of the SM and gives us a
testable prediction for one of them.  In fact, LEP data \cite{abf} 
seem to
suggest that a further symmetry, namely $N=1$ global supersymmetry
\cite{sakai1} 
should also be required to make the prediction viable.  Furthermore,
GUTs also relate Yukawa couplings among themselves, which in turn
might lead to testable predictions for the parameters of the SM.  The
prediction of the ratio $m_{\tau}/m_b$ \cite{begn} in the minimal
$SU(5)$ was an 
example of a successful reduction of the independent parameters of
this sector of the theory. On the other hand, requiring more symmetry
({\em e.g} $SO(10),~E_6,~E_7,~E_8$) does not necessarily lead to more
predictions for the SM parameters, due to the presence of new degrees
of freedom, various ways and channels of breaking the theory, etc.  An
extreme case from this point of view are superstrings, which have huge
symmetries, but no predictions for the SM parameters.

In a series of papers \cite{kmz,mondragon2,kmtz,kmoz} we have proposed
that a natural gradual 
extension of the GUTs ideas, which preserves their successes and
enhances the predictions, is to attempt to relate the gauge and Yukawa
couplings of a GUT, or in other words, to achieve Gauge-Yukawa
Unification (GYU).  
 
Searching for a symmetry that could provide such a
unification, one is led to introduce a symmetry that relates fields
with different spins, {\em i.e.} supersymmetry, and in particular
$N=2$ supersymmetry \cite{f79}.  
Unfortunately $N=2$ supersymmetric theories have
serious phenomenological problems due to light mirror fermions.
We expect that a GYU is a functional relationship which is derived by
some principle.  In superstring theories or in composite models 
there exist relations among gauge and Yukawa couplings, although 
in practice both kinds of theories have more problems than the SM.  

Before turning to our attempts, let us mention some earlier ones in
order to make clear which are really the predictions in each case.
Veltman, Decker and Pestieau \cite{depe}, by requiring the {\em absence of
  quadratic divergences} in the SM, found that the following
relationship has to hold:
\bea
m_e^2+m_{\mu}^2+m_{\tau}^2+3(m_u^2+m_d^2+m_c^2+m_s^2+m_t^2+m_b^2)&\nn\\
={3\over 2}m_W^2+{3\over 4}m_Z^2+{3\over 4}m_H^2 &
\label{uno}
\eea
A very similar relation is obtained demanding spontaneous breaking of
supersymmetry via F-terms.  In that case one obtains \cite{fgp}
\be
\sum_j (-1)^{2j} (2j+1)m_j^2 = 0,
\ee
where $j$ is the spin of the particle.  When this formula is applied
to the SM we obtain a relation 
which differs from Eq.(\ref{uno}) only in the coefficient of $m_H^2$.
In both cases a prediction for the top quark was possible only when it
was permitted experimentally to neglect the $m_H$ as compared to
$m_{W,Z}$ with the result $m_t=69$ GeV.  Otherwise there is only a
quadratic relation among $m_t$ and $m_H$.

A celebrated relation among gauge and Yukawa couplings is the
Pendleton-Ross (P-R) {\em infrared} fixed point \cite{pr}.  The P-R
proposal, involving the Yukawa coupling of the top quark $g_t$
and the strong gauge coupling $\alpha_3$, was that the ratio
$\alpha_t/\alpha_3$, where $\alpha_t=g_t^2/4\pi$, has an infrared fixed point.
This assumption predicted
$m_t \sim 100$ GeV, and therefore it is ruled out.  Moreover, it has
been shown \cite{zim-pr} that the P-R conjecture is not justified at
two-loops.  On the contrary, the ratio $\alpha_t/ \alpha _3$ diverges
in the infrared.

Another interesting conjecture, made by Hill \cite{hill},
is that  $\alpha _t$ itself develops an
infrared fixed point, leading to the prediction $m_t \sim 280$ GeV.

The P-R and Hill conjectures have been done in the framework on the SM.
The same conjectures within the minimal supersymmetric SM (MSSM) lead
to the following relations:
\bea
m_t &\simeq & 140 ~{\rm GeV}~\sin \beta ~~~(P-R)\label{3}\\
m_t &\simeq & 200 ~{\rm GeV}~\sin \beta ~~~(Hill)\label{4}
\eea
where $\tan \beta = {v_u/ v_d}$ is the ratio of the two VEV of the
Higgs fields of the MSSM.  We should stress that in  this case there
is no prediction for $m_t$, given that $\sin \beta$ is not fixed from
other considerations.  Therefore, the conclusion is that all the
attempts that have been made so far to extract predictions by some principle,
leading to relations among gauge and Yukawa couplings were found either
wrong or not  predictive enough.

In the following we would like to emphasize an alternative way to
achieve unification of couplings, which is based on the fact that
within the framework of a renormalizable field theory, one can find
renormalization group invariant (RGI) relations among parameters,
that can improve the calculability and the predictive power of a
theory.  In our recent studies \cite{kmz,mondragon2,kmtz,kmoz}, we have
considered the GYU 
which is based on the principles of reduction of couplings
\cite{cheng1,Zim,kubo1,kubo2,kubo3} and
finiteness \cite{kmz,PW,HPS,model1,model,al,LPS,pisi}.  These principles,
which are formulated in 
perturbation theory, are not explicit symmetry principles, although
they might imply symmetries.  The former principle is based on the
existence of RGI relations among couplings, which preserve
perturbative renormalizability.  Similarly, the latter one is based on
the fact that it is possible to find RGI relations among couplings
that keep finiteness in perturbation theory, even to all orders.
Applying these principles one can relate the gauge and Yukawa
couplings without introducing necessarily a symmetry, nevertheless
improving the predictive power of a model.

\section{ Unification of Couplings by the RGI Method}

Let us next briefly outline the idea.  
Any RGI relation among couplings 
(which does not depend on the renormalization
scale $\mu$ explicitly) can be expressed,
in the implicit form $\Phi (g_1,\cdots,g_A) ~=~\mbox{const.}$,
which
has to satisfy the partial differential equation (PDE)
\bea
\mu\,\frac{d \Phi}{d \mu} &=& {\vec \nabla}\cdot {\vec \beta} ~=~ 
\sum_{a=1}^{A} 
\,\beta_{a}\,\frac{\partial \Phi}{\partial g_{a}}~=~0~,
\eea
where $\beta_a$ is the $\beta$-function of $g_a$.
This PDE is equivalent
to the set  to ordinary differential equations, 
the so-called reduction equations (REs) \cite{Zim},
\bea
\beta_{g} \,\frac{d g_{a}}{d g} &=&\beta_{a}~,~a=1,\cdots,A~,
\label{redeq}
\eea
where $g$ and $\beta_{g}$ are the primary 
coupling and its $\beta$-function,
and $a$ does not include it.
Since maximally ($A-1$) independent 
RGI ``constraints'' 
in the $A$-dimensional space of couplings
can be imposed by $\Phi_a$'s, one could in principle
express all the couplings in terms of 
a single coupling $g$.
 The strongest requirement is to demand
 power series solutions to the REs,
\bea
g_{a} &=& \sum_{n=0}\rho_{a}^{(n+1)}\,g^{2n+1}~,
\label{powerser}
\eea
which formally preserve perturbative renormalizability.
Remarkably, the 
uniqueness of such power series solutions
can be decided already at the one-loop level \cite{Zim}.
To illustrate this, let us assume that the $\beta$-functions
have the form
\bea
\beta_{a} &=&\frac{1}{16 \pi^2}[
\sum_{b,c,d\neq g}\beta^{(1)\,bcd}_{a}g_b g_c g_d+
\sum_{b\neq g}\beta^{(1)\,b}_{a}g_b g^2]+\cdots~,\nn\\
\beta_{g} &=&\frac{1}{16 \pi^2}\beta^{(1)}_{g}g^3+
\cdots~,
\eea
where $\cdots$ stands for higher order terms, and 
$ \beta^{(1)\,bcd}_{a}$'s are symmetric in $
b,c,d$.
 We then assume that
the $\rho_{a}^{(n)}$'s with $n\leq r$
have been uniquely determined. To obtain $\rho_{a}^{(r+1)}$'s,
we insert the power series (\ref{powerser}) into the REs (\ref{redeq})
and collect terms of  
$O(g^{2r+3})$ and find
\bea
\sum_{d\neq g}M(r)_{a}^{d}\,\rho_{d}^{(r+1)} &=&
\mbox{lower order quantities}~,\nn
\eea
where the r.h.s. is known by assumption, and
\bea
M(r)_{a}^{d} &=&3\sum_{b,c\neq g}\,\beta^{(1)\,bcd}_{a}\,\rho_{b}^{(1)}\,
\rho_{c}^{(1)}+\beta^{(1)\,d}_{a}
-(2r+1)\,\beta^{(1)}_{g}\,\delta_{a}^{d}~,\label{M}\\
0 &=&\sum_{b,c,d\neq g}\,\beta^{(1)\,bcd}_{a}\,
\rho_{b}^{(1)}\,\rho_{c}^{(1)}\,\rho_{d}^{(1)}
+\sum_{d\neq g}\beta^{(1)\,d}_{a}\,\rho_{d}^{(1)}
-\beta^{(1)}_{g}\,\rho_{a}^{(1)}~.
\eea
 Therefore,
the $\rho_{a}^{(n)}$'s for all $n > 1$
for a given set of $\rho_{a}^{(1)}$'s can be uniquely determined if
$\det M(n)_{a}^{d} \neq 0$  for all $n \geq 0$.

The possibility of coupling unification described above 
is without any doubt
attractive because the ``completely reduced'' theory contains 
only one independent coupling, but  it can be
unrealistic. Therefore, one often would like to impose fewer RGI
constraints, and this is the idea of partial reduction \cite{kubo1}.

In the following chapters we would like to consider three different GYU
models based on supersymmetric unified theories.  Before doing so let
us recall here some features which are common in all cases.
  Let us consider a chiral, anomaly free,
$N=1$ globally supersymmetric
gauge theory based on a group G with the gauge coupling
constant $g$. The
superpotential of the theory is given by
\bea
W&=& \frac{1}{2}\,m_{ij} \,\phi_{i}\,\phi_{j}+
\frac{1}{6}\,C_{ijk} \,\phi_{i}\,\phi_{j}\,\phi_{k}~,
\label{supot}
\eea
where $m_{ij}$ and $C_{ijk}$ are gauge invariant tensors and
the matter field $\phi_{i}$ transforms
according to the irreducible representation  $R_{i}$
of the gauge group $G$. The
renormalization constants associated with the
superpotential (\ref{supot}), assuming that
supersymmetry is preserved, are
\bea
\phi_{i}^{0}&=&(Z^{j}_{i})^{(1/2)}\,\phi_{j}~,~\\
m_{ij}^{0}&=&Z^{i'j'}_{ij}\,m_{i'j'}~,~\\
C_{ijk}^{0}&=&Z^{i'j'k'}_{ijk}\,C_{i'j'k'}~.
\eea
The $N=1$ non-renormalization theorem \cite{nonre} ensures that
there are no mass
and cubic-interaction-term infinities and therefore
\bea
Z_{ijk}^{i'j'k'}\,Z^{1/2\,i''}_{i'}\,Z^{1/2\,j''}_{j'}
\,Z^{1/2\,k''}_{k'}&=&\delta_{(i}^{i''}
\,\delta_{j}^{j''}\delta_{k)}^{k''}~,\nn\\
Z_{ij}^{i'j'}\,Z^{1/2\,i''}_{i'}\,Z^{1/2\,j''}_{j'}
&=&\delta_{(i}^{i''}
\,\delta_{j)}^{j''}~.
\eea
As a result the only surviving possible infinities are
the wave-function renormalization constants
$Z^{j}_{i}$, i.e.,  one infinity
for each field. The one -loop $\beta$-function of the gauge
coupling $g$ is given by \cite{PW}
\bea
\beta^{(1)}_{g}=\frac{d g}{d t} =
\frac{g^3}{16\pi^2}\,[\,\sum_{i}\,l(R_{i})-3\,C_{2}(G)\,]~,
\label{betag}
\eea
where $l(R_{i})$ is the Dynkin index of $R_{i}$ and $C_{2}(G)$
 is the
quadratic Casimir of the adjoint representation of the
gauge group $G$. The $\beta$-functions of
$C_{ijk}$,
by virtue of the non-renormalization theorem, are related to the
anomalous dimension matrix $\gamma_{ij}$ of the matter fields
$\phi_{i}$ as:
\be
\beta_{ijk} =
 \frac{d C_{ijk}}{d t}~=~C_{ijl}\,\gamma^{l}_{k}+
 C_{ikl}\,\gamma^{l}_{j}+
 C_{jkl}\,\gamma^{l}_{i}~.
\label{betay}
\ee
At one-loop level $\gamma_{ij}$ is \cite{PW}
\be
\gamma_{ij}^{(1)}=\frac{1}{32\pi^2}\,[\,
C^{ikl}\,C_{jkl}-2\,g^2\,C_{2}(R_{i})\delta_{ij}\,],
\label{gamay}
\ee
where $C_{2}(R_{i})$ is the quadratic Casimir of the representation
$R_{i}$, and $C^{ijk}=C_{ijk}^{*}$.
Since
dimensional coupling parameters such as masses  and couplings of cubic
scalar field terms do not influence the asymptotic properties 
 of a theory on which we are interested here, it is
sufficient to take into account only the dimensionless supersymmetric
couplings such as $g$ and $C_{ijk}$.
So we neglect the existence of dimensional parameters, and
assume furthermore that
$C_{ijk}$ are real so that $C_{ijk}^2$ always are positive numbers.
For our purposes, it is
convenient to work with the square of the couplings and to
arrange $C_{ijk}$ in such
a way that they are covered by a single index $i~(i=1,\cdots,n)$:
\bea
\alpha &=& \frac{|g|^2}{4\pi}~,~
\alpha_{i} ~=~ \frac{|g_i|^2}{4\pi}~.
\label{alfas}
\eea

The evolution equations of $\alpha$'s in perturbation theory
then take the
form
 \bea
\frac{d\alpha}{d t}&=&\beta~=~ -\beta^{(1)}\alpha^2+\cdots~,\nn\\
\frac{d\alpha_{i}}{d t}&=&\beta_{i}~=~ -\beta^{(1)}_{i}\,\alpha_{i}\,
\alpha+\sum_{j,k}\,\beta^{(1)}_{i,jk}\,\alpha_{j}\,
\alpha_{k}+\cdots~,
\label{eveq}
\eea
where
$\cdots$ denotes the contributions from higher orders, and
$ \beta^{(1)}_{i,jk}=\beta^{(1)}_{i,kj}  $.

Given the set of the evolution equations (\ref{eveq}), we investigate the
asymptotic  properties, as follows. First we
 define \cite{cheng1,Zim}
\bea
\tilde{\alpha}_{i} &\equiv& \frac{\alpha_{i}}{\alpha}~,~i=1,\cdots,n~,
\label{alfat}
\eea
and derive from Eq. (\ref{eveq})
\bea
\alpha \frac{d \tilde{\alpha}_{i}}{d\alpha} &=&
-\tilde{\alpha}_{i}+\frac{\beta_{i}}{\beta}~=~
(\,-1+\frac{\beta^{(1)}_{i}}{\beta^{(1)}}\,)\, \tilde{\alpha}_{i}\nn\\
& &
-\sum_{j,k}\,\frac{\beta^{(1)}_{i,jk}}{\beta^{(1)}}
\,\tilde{\alpha}_{j}\, \tilde{\alpha}_{k}+\sum_{r=2}\,
(\frac{\alpha}{\pi})^{r-1}\,\tilde{\beta}^{(r)}_{i}(\tilde{\alpha})~,
\label{RE}
\eea
where $\tilde{\beta}^{(r)}_{i}(\tilde{\alpha})~(r=2,\cdots)$
are power series of $\tilde{\alpha}$'s and can be computed
from the $r$-th loop $\beta$-functions.
Next we search for fixed points $\rho_{i}$ of Eq. (\ref{alfat}) at $ \alpha
= 0$. To this end, we have to solve
\bea
(\,-1+\frac{\beta ^{(1)}_{i}}{\beta ^{(1)}}\,)\, \rho_{i}
-\sum_{j,k}\frac{\beta ^{(1)}_{i,jk}}{\beta ^{(1)}}
\,\rho_{j}\, \rho_{k}&=&0~,
\label{fixpt}
\eea
and assume that the fixed points have the form
\bea
\rho_{i}&=&0~\mbox{for}~ i=1,\cdots,n'~;~
\rho_{i} ~>0 ~\mbox{for}~i=n'+1,\cdots,n~.
\eea
We then regard $\tilde{\alpha}_{i}$ with $i \leq n'$
 as small
perturbations  to the
undisturbed system which is defined by setting
$\tilde{\alpha}_{i}$  with $i \leq n'$ equal to zero.
As we have seen,
it is possible to verify at the one-loop level \cite{Zim} the
existence of the unique power series solution
\bea
\tilde{\alpha}_{i}&=&\rho_{i}+\sum_{r=2}\rho^{(r)}_{i}\,
\alpha^{r-1}~,~i=n'+1,\cdots,n~
\label{usol}
\eea
of the reduction equations (\ref{RE}) to all orders in the undisturbed
system. 
These are RGI relations among couplings and keep formally
perturbative renormalizability of the undisturbed system.
So in the undisturbed system there is only {\em one independent}
coupling, the primary coupling $\alpha$.

 The small
 perturbations caused by nonvanishing $\tilde{\alpha}_{i}$
 with $i \leq n'$
enter in such a way that the reduced couplings,
i.e., $\tilde{\alpha}_{i}$  with $i > n'$,
become functions not only of
$\alpha$ but also of $\tilde{\alpha}_{i}$
 with $i \leq n'$.
It turned out that, to investigate such partially
reduced systems, it is most convenient to work with the partial
differential equations
\bea
\{~~\tilde{\beta}\,\frac{\partial}{\partial\alpha}
+\sum_{a=1}^{n'}\,
\tilde{\beta_{a}}\,\frac{\partial}{\partial\tilde{\alpha}_{a}}~~\}~
\tilde{\alpha}_{i}(\alpha,\tilde{\alpha})
&=&\tilde{\beta}_{i}(\alpha,\tilde{\alpha})~,\nn\\
\tilde{\beta}_{i(a)}~=~\frac{\beta_{i(a)}}{\alpha^2}
-\frac{\beta}{\alpha^{2}}~\tilde{\alpha}_{i(a)}
&,&
\tilde{\beta}~\equiv~\frac{\beta}{\alpha}~,
\eea
 which are equivalent
to the reduction equations (\ref{RE}), where we let
$a,b$ run from $1$ to $n'$ and $i,j$ from $n'+1$ to $n$
in order to avoid confusion.
We then look for solutions of the form
\bea
\tilde{\alpha}_{i}&=&\rho_{i}+
\sum_{r=2}\,(\frac{\alpha}{\pi})^{r-1}\,f^{(r)}_{i}
(\tilde{\alpha}_{a})~,~i=n'+1,\cdots,n~,
\label{algeq}
\eea
where $ f^{(r)}_{i}(\tilde{\alpha}_{a})$ are supposed to be
power series of
$\tilde{\alpha}_{a}$. This particular type of solution
can be motivated by requiring that in the limit of vanishing
perturbations we obtain the undisturbed
solutions (\ref{usol}) \cite{kubo2,zimmermann3}.
Again it is possible to obtain  the sufficient conditions for
the uniqueness of $ f^{(r)}_{i}$ in terms of the lowest order
coefficients.

\section{Finite Unified Model Based on $SU(5)$}

According to the discussion in the previous chapter, the
 non-renormalization theorem ensures that
there are no extra mass
and cubic-interaction-term renormalizations, implying that
the $\beta$-functions of
$C_{ijk}$ can be expressed as linear combinations of the
anomalous dimension matrix $\gamma_{ij}$ of
$\phi^{i}$.
Therefore, all the one-loop $\beta$-functions of the theory vanish
if
\bea
\beta_{g}^{(1)}&=&0~\mbox{and}~\gamma_{ij}^{(1)}~=~0~
\label{betagama}
\eea
are satisfied,
where $\beta_{g}^{(1)}$  and $\gamma_{ij}^{(1)}$
are given in Eqs. (\ref{betag}) and (\ref{gamay}) respectively.
A very interesting result is that these conditions (\ref{betagama}) are
necessary and sufficient for finiteness at
the two-loop level \cite{PW}.

A natural question is what happens in higher loops.
Since the finiteness conditions impose relations
among couplings, they have to be consistent with the
REs (\ref{redeq}) (this should be so even for the one-loop finiteness).
Interestingly,
 there exists a powerful theorem \cite{LPS}
which provides the necessary and sufficient conditions for
finiteness to all loops.
The theorem makes heavy use of the non-renormalization
property of the supercurrent anomaly \cite{pisi}.
In fact, the  finiteness theorem can be formulated in terms of
one-loop quantities, and it states
that for supersymmetry gauge theories we are considering here, the necessary
and sufficient conditions for $\beta_{g}$ and $\beta_{ijk}$ to
vanish to all orders are \cite{LPS}: \newline
(a) The validity of the one-loop finiteness conditions, i.e.,
 Eq. (\ref{betagama}) is satisfied.
\newline
(b) The REs (2) admit a unique power series
solution, i.e., the corresponding matrix $M$ defined in Eq. (\ref{M}) with
$\beta_{g}^{(1)}=0$ has to be non-singular.
\newline
The latter condition
is equivalent to the requirement that the one-loop
solutions
$\rho_{a}^{(1)}$'s
are isolated and non-degenerate. Then each of these solutions
can be extended, by a recursion formula, to a formal power series in $g$
giving a theory which depends on a single coupling $g$, and has
$\beta$-functions vanishing to all orders.

{}From the classification of
theories with $\beta_{g}^{(1)}=0$
\cite{HPS}, one can see that
using $SU(5)$ as gauge group there
exist only two candidate models which can accommodate three fermion
generations. These models contain the chiral supermutiplets
${\bf 5}~,~\overline{\bf 5}~,~{\bf 10}~,~\overline{\bf 5}~,~{\bf 24}$
with the multiplicities $(6,9,4,1,0)$ and
 $(4,7,3,0,1)$, respectively.
Only the second one contains a ${\bf 24}$-plet which can be used
for spontaneous symmetry breaking (SSB) of $SU(5)$ down
to $SU(3)\times SU(2) \times U(1)$. (For the first model
one has to incorporate another way, such as the Wilson flux
breaking to achieve the desired SSB of $SU(5)$.)
Here we would like to concentrate only on the second model.

The most general $SU(5)$ invariant, cubic
superpotential of the (second)
model is:
\bea
W &=&H_{a}\,[~
f_{ab}\,\overline{H}_b {\bf 24}+
h_{ia}\,\overline{\bf 5}_i {\bf 24}
+\overline{g}_{ija}\,{\bf 10}_i \overline{\bf 5}_{j}]+
 p\,({\bf 24})^3 \nn\\
&+& \frac{1}{2}\,{\bf 10}_i\,[~
g_{ija}\,{\bf 10}_j H_a+
 \hat{g}_{iab}\,\overline{H}_a
\overline{H}_b+
g_{ijk}^{\prime}\,
\overline{\bf 5}_{j} \overline{\bf 5}_{k}~]~,
\eea
where $i,j,k=1,2,3$ and $a,b=1,\cdots,4$, and we sum over all
indices  in $W$ (the $SU(5)$ indices are suppressed).
The ${\bf 10}_{i}$'s
and $\overline{\bf 5}_{i}$'s are the usual
three generations, and the four
$({\bf 5}+ \overline{\bf 5})$ Higgses are denoted by
 $H_a~,~\overline{H}_{a} $.

Given the superpotential, the
$\gamma^{(1)}$'s can be easily computed ($\beta_{g}^{(1)}$
vanishes of course). To ensure finiteness of the model
to all orders, we have to find
$\rho^{(1)}$'s that are isolated and non-degenerate solutions
of Eq. (6) and
are consistent with the vanishing $\gamma^{(1)}$'s.
In most of the previous studies of
the present model \cite{model1,model}, however,
no attempt was made to find isolated and non-degenerate
solutions, but rather the opposite. They have used the freedom
offered by the degeneracy in order to make specific ans{\" a}tze
that could lead to phenomenologically acceptable
predictions. Here we concentrate on finding an
isolated and non-degenerate solution that is phenomenologically
interesting. As a first approximation to the Yukawa
matrices, a diagonal solution, that is, without
intergenerational mixing, may be considered.
It has turned out that this can be achieved by imposing
the $Z_{7}\times Z_{3}$ discrete symmetry and
a multiplicative $Q$-parity  on $W$, and that, in order
to respect these symmetries,
only $ g_{iii}~,~\overline{g}_{iii}~,
~f_{ii}$ and $p$ are allowed to be non-vanishing.
Moreover, we have found that under this situation
there exists a unique reduction solution that satisfies the finiteness
conditions (a) and (b) \cite{kmz}:
\bea
\alpha_{iii}&=&\frac{8}{5}\alpha_{GUT}+O(\alpha_{GUT}^2)~,~
\overline{\alpha}_{iii}~=~\frac{6}{5}\alpha_{GUT}+O(\alpha_{GUT}^2)~,~
 \alpha_{f_{ii}}~=~0~,\nn\\
\alpha_ {f_{44}}&=&\alpha_{GUT}+O(\alpha_{GUT}^2)~,
~\alpha_p~=~\frac{15}{7}\alpha_{GUT}+O(\alpha_{GUT}^2)~,
\label{solfut}
\eea
where $i=1,2,3$, and
the $O(\alpha_{GUT}^2)$ terms are power series in $\alpha_{GUT}$ that can be
uniquely 
computed to any finite order if the $\beta$-functions
of the unreduced model are known to the corresponding order.
The reduced model in which gauge and Yukawa couplings
are unified
has the $\beta$-functions that identically vanish to that order.

In the above model, we found a diagonal solution for the Yukawa
couplings, with each family coupled to a different Higgs.
However, we may use the fact that mass terms
do not influence the $\beta$-functions in a certain
class of renormalization schemes, and introduce
appropriate mass terms that permit us to perform a rotation in the Higgs
sector such that only one pair of Higgs doublets, coupled to
the third family, remains light and acquires a
non-vanishing VEV (in a similar way to what was done by
Le\'on et al.~\cite{model}). 
Note that the effective coupling of the Higgs doublets
to the first family after
the rotation is very small avoiding in this way a potential problem
with the proton lifetime \cite{proton}.
Thus, effectively,
we have at low energies the MSSM with
only one pair of Higgs doublets. Adding soft
breaking terms (which are supposed not to influence the
$\beta$-functions beyond $M_{GUT}$),
we can obtain supersymmetry breaking.
The conditions on the soft breaking terms to preserve
one-loop finiteness have been given already some time ago
\cite{soft}. 
Recently, the same problem
at the two-loop level has been addressed \cite{jones}.
It is an open problem whether there exists a suitable set of conditions
on the soft terms for all-loop finiteness.
Since the $SU(5)$ symmetry is spontaneously broken
below $M_{GUT}$, the finiteness conditions obviously
do not restrict the renormalization property at low energies, and
all it remains is a boundary condition on the
gauge and Yukawa couplings; these couplings at low energies
have to be so chosen that they satisfy  (\ref{solfut}) at $M_{GUT}$.
So we examine the evolution of the gauge and Yukawa couplings according
to their renormalization group equations at two-loops taking into
account all the boundary conditions at $M_{GUT}$.
The predictions for $m_t$ for a varying, but common to all particles,
supersymmetry breaking 
threshold are given in Figure 1.

\section{The Minimal Asymptotically Free SU(5) Model}

Let us consider next the {\em minimal} $N=1$ supersymmetric gauge model
based on the group $SU(5)$ \cite{sakai1}. Its particle content is then
specified and has the following transformation properties under $SU(5)$:
three $({\bf \overline{5}}+{\bf 10})$-
supermultiplets which accommodate three fermion families,
one $({\bf 5}+{\bf \overline{5}})$ to describe the two Higgs
supermultiplets appropriate for electroweak symmetry breaking
and a ${\bf 24}$-supermultiplet required to provide the
spontaneous symmetry breaking of $SU(5)$ down to
$SU(3)\times SU(2) \times U(1)$.

Since we are neglecting the dimensional parameters
and the Yukawa couplings of the first two generations,
the
superpotential of the model is exactly given by
\begin{equation}
W = \frac{1}{2}\,\,g_{t} {\bf 10}_{3}\,
{\bf 10}_{3}\,H+
g_{b}\, \overline{{\bf 5}}_{3}\,{\bf 10}_{3}\, \overline{H}
+g_{\lambda}\,({\bf 24})^3+
g_{f}\,\overline{H}\,{\bf 24}\, H~,
\end{equation}
where $H, \overline{H}$ are the ${\bf 5},\overline{{\bf 5}}$-
Higgs supermultiplets and we have suppressed the $SU(5)$
indices.
According to the notation introduced in Eq.~(\ref{alfat}), 
Eqs.~(\ref{RE}) become
\bea
\alpha\,\frac{d \tilde{\alpha}_{t}}{d \alpha} &=&
\frac{27}{5}\,\tilde{\alpha}_{t}
-3\,\tilde{\alpha}_{t}^2-\frac{4}{3}\,\tilde{\alpha}_{t}
\tilde{\alpha}_{b}-
\frac{8}{5}\,\tilde{\alpha}_{t}\,\tilde{\alpha}_{f}~,\nn\\
\alpha\,\frac{d \tilde{\alpha}_{b}}{d \alpha} &=&
\frac{23}{5}\,\tilde{\alpha}_{b}
-\frac{10}{3}\,\tilde{\alpha}_{b}^2-\tilde{\alpha}_{b}
\tilde{\alpha}_{t}-
\frac{8}{5}\,\tilde{\alpha}_{b}\,\tilde{\alpha}_{f}~,\nn\\
\alpha\,\frac{d \tilde{\alpha}_{\lambda}}{d \alpha} &=&
9\tilde{\alpha}_{\lambda} -\frac{21}{5}\,\tilde{\alpha}_{\lambda}^2-
\tilde{\alpha}_{\lambda}\,\tilde{\alpha}_{f}~,\nn\\ \alpha\,\frac{d
\tilde{\alpha}_{f}}{d \alpha} &=&
\frac{83}{15}\,\tilde{\alpha}_{f}
-\frac{53}{15}\,\tilde{\alpha}_{f}^2-\tilde{\alpha}_{f}
\tilde{\alpha}_{t}-
\frac{4}{3}\,\tilde{\alpha}_{f}\,\tilde{\alpha}_{b}-
\frac{7}{5}\,\tilde{\alpha}_{f}\,\tilde{\alpha}_{\lambda}~,
\eea
in the one-loop approximation.
Given the above equations describing the evolution of the four
independent couplings $(\alpha_{i}~,~i=t,b,\lambda,f)$,
there exist $2^4=16$ non-degenerate solutions corresponding
to vanishing $\rho$'s as well as non-vanishing ones
given by Eq.~(\ref{algeq})
The
possibility to predict the top quark mass depends, as in the previous
model,  on a nontrivial
interplay between the vacuum expectation value of the two $SU(2)$
Higgs
doublets involved in the model and the known masses of the third
generation $(m_{b}~,~m_{\tau})$.
It is clear that only the solutions of the form
\be
\rho_{t}~,~\rho_{b}~\neq~0
\label{33}
\ee
can predict the top and bottom quark masses.

There exist exactly  four such solutions.
The first solution is ruled out since it is inconsistent
with Eq.~(\ref{alfas}), and the second one is ruled out
since it does not 
satisfy the criteria to be asymptotically free.
We are left with two asymptotically free solutions, 
which we label 3 and 4.
According to the criteria of section 2, 
these two solutions
give the possibility to obtain  partial reductions.
To achieve this, we look for solutions \cite{mondragon2} of the form 
Eq.~(\ref{usol}) to both 3 and 4.

We present now the computation of some lower order terms within the
one-loop approximation for the solutions.  For solution 3:
\bea
\tilde{\alpha}_{i} &=& \eta_{i}+ f^{(r_{\lambda}=1)}
_{i}\,\tilde{\alpha}_{\lambda}+f^{(r_{\lambda}=2)}
_{i}\,\tilde{\alpha}_{\lambda}^2+\cdots~~\mbox{for}~i=t,b,f~,
\label{39}
\eea
where
\bea
\eta_{t,b,f} &=&\frac{2533}{2605}~,~
\frac{1491}{2605}~,~\frac{560}{521}~,\nn\\
f^{(r_{\lambda}=1)}_{t,b,f}&\simeq & 0.018~,~0.012~,~-0.131~,\nn\\
f^{(r_{\lambda}=2)}_{t,b,f} &\simeq & 0.005~,~0.004~,~-0.021~,
\label{40}
\eea
For the solution $4$,
\bea
\tilde{\alpha}_{i} &=& \eta_{i}+f^{(r_{f}=1)}_{i}\,
\tilde{\alpha}_{f}+f^{(r_{\lambda}=1)}_{i}\,
\tilde{\alpha}_{\lambda}+f^{(r_{f}=1,r_{\lambda}=1)}_{i}\,
\tilde{\alpha}_{f}\,\tilde{\alpha}_{\lambda}\nn\\
& &+f^{(r_{f}=2)}_{i}\,
\tilde{\alpha}_{f}^2+f^{(r_{\lambda}=2)}_{i}\,
\tilde{\alpha}_{\lambda}^{2}\cdots~~\mbox{for}~i=t,b~,
\label{41}
\eea
where
\bea
 \eta_{t,b}&=&\frac{89}{65}
~,~\frac{63}{65}~,~f^{(r_{\lambda}=1)}_{i}~=~
f^{(r_{\lambda}=2)}_{i}~=~0~,\nn\\
f^{(r_{f}=1)}_{t,b}
&\simeq & -0.258~,
~-0.213~,~f^{(r_{f}=1)}_{t,b}
~\simeq ~ -0.258~,
~-0.213~,\nn\\
f^{(r_{f}=2)}_{t,b}
& \simeq & -0.055~,~ -0.050~,~
f^{(r_{f}=1,r_{\lambda}=1)}_{t,b}
~\simeq ~ -0.021~,~-0.018 ~,
\label{42}
\eea
In the solutions (\ref{39}) and (\ref{41}) we have 
suppressed the contributions
from the Yukawa couplings of the first two generations
because they are negligibly small.

Presumably, both solutions are related;
a numerical analysis on the solutions \cite{mondragon2} suggests that
the solution 
$3$ is a ``boundary''
of $4$. If it is really so,
then there is only one unique reduction solution in the minimal
supersymmetric GUT that provides us with  the possibility of predicting
$\alpha_{t}$. Note furthermore that not only
$\alpha_{t}$ but also $\alpha_{b}$
is predicted in this reduction solution.

Just below the unification scale we would like to obtain the standard
$SU(3)\times SU(2)\times U(1)$ model while assuming that
all the superpartners are decoupled at the Fermi scale.
Then the standard model should  be spontaneously broken down to
$SU(3)\times U(1)_{\rm em}$ due to VEV of the two Higgs
$SU(2)$-doublets contained in the ${\bf 5},\overline{{\bf
5}}$-super-multiplets.
One way to obtain the correct low energy theory is to add to
the Lagrangian soft supersymmetry breaking
terms and to arrange
the mass parameters in the superpotential along with
the soft breaking terms so that
the desired symmetry breaking pattern of the original $SU(5)$
is really the preferred one, all the superpartners are
unobservable at present energies,
there is no contradiction with proton decay,
and so forth.
Then, as in the previous model, we study 
the evolution of the couplings at two loops
respecting all the boundary conditions at $M_{GUT}$.
The predictions for $m_t$ versus $M_{SUSY}$ for the two sets of
boundary conditions given above (AFUT3 and AFUT4) together with the
corresponding predictions of the FUT model, are given in Figure 1.
In a  recent study \cite{KYIS95}, we have 
considered the proton decay constraint \cite{HMY-npb402}
to further reduce the parameter space of the model.
It has been found that the model consistent with the 
non-observation of the proton decay
should be very close to AFUT3, implying a better
possibility to discriminate between the FUT and AFUT models,
as one can see from Figure 4.

\section{ Asymptotically Non-Free Supersymmetric Pati-Salam Model}

In order for the RGI method for the gauge coupling
unification to  work,
the gauge couplings should 
have the same asymptotic behavior.
Note that this common behavior is absent
in the standard model with three families.
A way to achieve a common asymptotic behavior of all the
different gauge couplings is to embed 
$SU(3)_{C}\times SU(2)_{L}\times U(1)_{Y}$ to some
non-abelian gauge group, as it was done in sections 3 and 4.
However, in this case still a major role in the
GYU is due to the group theoretical aspects of the covering GUT. Here
we would like  to examine the power of RGI method by considering
theories without covering GUTs.
We found \cite{kmtz}
that the minimal phenomenologically viable model is based on the gauge
group of Pati and 
Salam \cite{pati1}-- ${\cal G}_{\rm PS}\equiv 
SU(4)\times SU(2)_{R}\times
SU(2)_{L}$.
We recall that $N=1$ supersymmetric  models based on this
gauge group have been studied with renewed interest because they could
in principle be derived from superstring \cite{anton1}.

In our supersymmetric, Gauge-Yukawa unified model
based on $ {\cal G}_{\rm PS}$ \cite{kmtz}, three generations of
quarks and leptons  are accommodated by six chiral supermultiplets, three
in $({\bf 4},{\bf 2},{\bf 1})$ and three  $({\bf \overline{4}},{\bf
1},{\bf 2})$, which we denote by $\Psi^{(I)\mu~ i_R}$ and $
\overline{\Psi}_{\mu}^{(I) i_L}$. ($I$ runs over the three generations,
and
$\mu,\nu~(=1,2,3,4)$ are the $SU(4)$ indices while 
$i_R~,~i_L~(=1,2)$ 
stand for the
$SU(2)_{L,R}$ indices.) 
The Higgs supermultiplets 
in $({\bf 4},{\bf 2},{\bf 1})$,
$({\bf \overline{4}},{\bf 2},{\bf 1})$
and  $({\bf 15},{\bf 1},{\bf 1})$ are denoted by 
$ H^{\mu ~i_R}~,~
\overline{H}_{\mu ~i_R} $ and $\Sigma^{\mu}_{\nu}$, respectively. They
 are responsible for the spontaneous
symmetry breaking (SSB) of $SU(4)\times SU(2)_{R}$ down 
to $SU(3)_{C}\times U(1)_{Y}$.
The SSB of $U(1)_{Y}\times
SU(2)_{L}$ is then achieved by the nonzero VEV of
$h_{i_R i_L}$ which is in $({\bf 1},{\bf 2},{\bf 2})$. In addition to
these Higgs supermultiplets, we introduce $G^{\mu}_{\nu~i_R i_L}~
({\bf 15},{\bf 2},{\bf 2})~,
~\phi~({\bf 1},{\bf 1},{\bf 1})$ and 
$\Sigma^{' \mu}_{\nu}~({\bf 15},{\bf 1},{\bf 1})$.
The $G^{\mu}_{\nu~i_R i_L}$ is introduced to realize 
the $SU(4)\times SU(2)_{R}\times
SU(2)_{L}$ version of the Georgi-Jarlskog type 
ansatz \cite{georgi4} for
the mass matrix of leptons and quarks while $\phi$ 
is supposed to mix with the right-handed neutrino
supermultiplets at a high energy scale.
With these in mind, we write down
the superpotential of the model
$W$, which is the sum of the following superpotentials:
\bea
W_{Y} &=&\sum_{I,J=1}^{3}g_{IJ}\,\overline{\Psi}^{(I) i_R}_{\mu} 
\,\Psi^{(J)\mu~ i_L}~h_{i_R i_L}~,~
W_{GJ} ~=~g_{GJ}\,
\overline{\Psi}^{(2)i_R}_{\mu}\,
G^{\mu}_{\nu~i_R j_L}\,\Psi^{(2)\nu~ j_L}~,\nn\\
W_{NM} &=&
\sum_{I=1,2,3}\,g_{I\phi}~\epsilon_{i_R j_R}\,\overline{\Psi}^{(I)
i_R}_{\mu} ~H^{\mu ~j_R}\,\phi~,\nn\\
W_{SB} &=&
g_{H}\,\overline{H}_{\mu~ i_R}\,
\Sigma^{\mu}_{\nu}\,H^{\nu ~i_R}+\frac{g_{\Sigma}}{3}\,
\mbox{Tr}~[~\Sigma^3~]+
\frac{g_{\Sigma '}}{2}\,\mbox{Tr}~[~(\Sigma ')^2\,\Sigma~]~,\nn\\
W_{TDS} &=&
\frac{g_{G}}{2}\,\epsilon^{i_R j_R}\epsilon^{i_L j_L}\,\mbox{Tr}~
[~G_{i_R i_L}\,
\Sigma\,G_{j_R j_L}~]~,\nn\\
W_{M}&=&m_{h}\,h^2+m_{G}\,G^2+m_{\phi}\,
\phi^2+m_{H}\,\overline{H}\,H+
m_{\Sigma}\,\Sigma^2+
m_{\Sigma '}\,(\Sigma ')^2~.\label{12} 
\eea
Although $W$ has the parity, $\phi\to -\phi$
and $\Sigma ' \to -\Sigma '$, 
it is not the most general potential, and,
by virtue of the non-renormalization theorem,
this does not contradict the philosophy of 
the coupling unification by the RGI method.

We denote the gauge couplings of $SU(4)\times SU(2)_{R}\times
SU(2)_{L}$
by $\alpha_{4}~,~\alpha_{2R}$ and $\alpha_{2L}$,
respectively. The gauge coupling for $U(1)_{Y}$, $\alpha_1$, normalized
in the usual GUT inspired manner, is given by
$1/\alpha_{1} ~=~2/5\alpha_{4}+
3/5 \alpha_{2R}~$.
In principle, the primary coupling can be any one of the couplings.
But it is more convenient to choose a gauge coupling as the primary
one because the one-loop $\beta$ functions for a gauge coupling
depends only on its own gauge coupling. For the present model,
we use $\alpha_{2L}$ as the primary one.
Since the gauge sector for the one-loop $\beta$ functions is closed,
the solutions of the fixed point equations (\ref{fixpt}) are 
independent on the Yukawa and Higgs couplings. One easily obtains
$
\rho_{4}^{(1)} =8/9~,~\rho_{2R}^{(1)}~=~4/5$,
so that the
RGI relations (\ref{algeq}) at the one-loop level become
\bea
\tilde{\alpha}_{4} &=&\frac{\alpha_4}{\alpha_{2L}}~=~
\frac{8}{9}~,~\tilde{\alpha}_{1} ~=~\frac{\alpha_1}{\alpha_{2L}}~=~
\frac{5}{6}~.
\label{13}
\eea

The solutions in the Yukawa-Higgs sector strongly
depend on the result of the gauge sector. After slightly involved
algebraic computations, one finds that 
most predictive solutions contain at least
three vanishing $\rho_{i}^{(1)}$'s.  
Out of these solutions, there are two that
 exhibit the most predictive
power and moreover they satisfy
the neutrino mass relation
$m_{\nu_{\tau}}~>~m_{\nu_{\mu}}~,~
m_{\nu_{e}}$. 
For the first solution we have $\rho_{1\phi}^{(1)}=
\rho_{2\phi}^{(1)}=
\rho_{\Sigma}^{(1)}=0$, while for the second solution, 
$  \rho_{1\phi}^{(1)}=
\rho_{2\phi}^{(1)}=
\rho_{G}^{(1)}=0 $,
 and one finds that for the cases above the power series solutions
(\ref{algeq}) take the form 
\bea
\tilde{\alpha}_{GJ} &\simeq &
\left\{
\begin{array}{l} 1.67 - 0.05 \tilde{\alpha}_{1\phi}
+ 
0.004 \tilde{\alpha}_{2\phi}
 - 0.90\tilde{\alpha}_{\Sigma}+\cdots \\
 2.20 - 0.08 \tilde{\alpha}_{2\phi}
 - 0.05\tilde{\alpha}_{G}+\cdots 
\end{array} \right. ~~,\nn\\
\tilde{\alpha}_{33} &\simeq&\left\{
\begin{array}{l}  3.33 + 0.05 \tilde{\alpha}_{1\phi} 
+ 
0.21 \tilde{\alpha}_{2\phi}-0.02 \tilde{\alpha}_{\Sigma}+ \cdots 
\\3.40 + 0.05 \tilde{\alpha}_{1\phi} 
-1.63 \tilde{\alpha}_{2\phi}- 0.001 \tilde{\alpha}_{G}+ 
\cdots \end{array} \right. ~~,\nn\\
\tilde{\alpha}_{3\phi} &\simeq&
\left\{
\begin{array}{l}  1.43 -0.58 \tilde{\alpha}_{1\phi} 
- 
1.43 \tilde{\alpha}_{2\phi}-0.03 \tilde{\alpha}_{\Sigma}+ 
\cdots\\
 0.88 -0.48 \tilde{\alpha}_{1\phi} 
+8.83 \tilde{\alpha}_{2\phi}+ 0.01 \tilde{\alpha}_{G}+ 
\cdots\end{array} \right. ~~,\nn\\
\tilde{\alpha}_{H} &\simeq& \left\{
\begin{array}{l}
 1.08 -0.03 \tilde{\alpha}_{1\phi} 
+0.10 \tilde{\alpha}_{2\phi}- 0.07 \tilde{\alpha}_{\Sigma}+ 
\cdots\nn\\
2.51 -0.04 \tilde{\alpha}_{1\phi} 
-1.68 \tilde{\alpha}_{2\phi}- 0.12 \tilde{\alpha}_{G}+ 
\cdots\end{array} \right. ~~,~~\\
\tilde{\alpha}_{\Sigma} &\simeq& \left\{
\begin{array}{l}
---\\
0.40 +0.01 \tilde{\alpha}_{1\phi} 
-0.45 \tilde{\alpha}_{2\phi}-0.10 \tilde{\alpha}_{G}+ 
\cdots \end{array} \right. ~,\nn\\  
\tilde{\alpha}_{\Sigma '} &\simeq& \left\{
\begin{array}{ll}
4.91 - 0.001 \tilde{\alpha}_{1\phi} 
-0.03 \tilde{\alpha}_{2\phi}- 0.46 \tilde{\alpha}_{\Sigma}+ 
\cdots
\\8.30 + 0.01 \tilde{\alpha}_{1\phi} 
+1.72 \tilde{\alpha}_{2\phi}- 0.36 \tilde{\alpha}_{G}+ 
\cdots \end{array} \right. ~~,  \nn\\
\tilde{\alpha}_{G} &\simeq& \left\{
\begin{array}{ll}
5.59 + 0.02 \tilde{\alpha}_{1\phi} 
-0.04 \tilde{\alpha}_{2\phi}- 1.33 \tilde{\alpha}_{\Sigma}+ 
\cdots
\\--- \end{array} \right.   ~ ~.\label{14}
\eea
We have assumed that the Yukawa couplings $g_{IJ}$ except for
$g_{33}$ vanish. They can be included into RGI relations
as small 
perturbations,
but their numerical effects
will be rather small.

 The number $ N_{H}$ of the Higgses lighter
than $M_{SUSY}$ could vary from one to four while the number of
those to be taken into account above $M_{SUSY}$ is fixed at four.
We have assumed here that $N_{H}=1$. The dependence of the top mass on
$M_{SUSY}$ in this model is shown in Figure 2.

\section{Discussion and Conclusions}

As a natural extension of the unification of gauge couplings provided by 
all GUTs and the model dependent unification of Yukawa couplings, we
have introduced in a number of publications the idea of Gauge-Yukawa
Unification. GYU is a functional relationship among the gauge and
Yukawa couplings provided by some principle.  In our studies GYU has
been achieved by demanding the principles of reduction of couplings
and finiteness.  The first principle is based on the existence of RGI
relations among couplings which preserve perturbative
renormalizability in gauge
theories.  The second principle is based on the fact that it is
possible to find RGI relations among couplings that keep finiteness in
perturbation theory, even to all orders.  In the previous chapters we
have presented the application of these principles in various models
as well as the resulting predictions.

The consequence of GYU is that 
in the lowest order in perturbation theory
 the gauge and Yukawa couplings above  $M_{\rm GUT}$
are related  in the form
\bea
g_i& = &\kappa_i \,g_{\rm GUT}~,~i=1,2,3,e,\cdots,\tau,b,t~,
\label{giki}
\eea
where $g_i~(i=1,\cdots,t)$ stand for the gauge 
and Yukawa couplings, $g_{\rm GUT}$ is the unified coupling,
and
we have neglected  the Cabibbo-Kobayashi-Maskawa mixing 
of the quarks.
 So, Eq. (\ref{giki}) exhibits a boundary condition on the 
the renormalization group evolution for the effective theory
below $M_{\rm GUT}$, which we assume to 
be the MSSM.

As we have seen in the previous chapters, there are various 
supersymmetric GUTs with GYU in the 
third generation that can
 predict the bottom and top
quark masses in accordance with the experimental data
\cite{kmz,mondragon2,kmtz}. 
This means that the top-bottom hierarchy 
could be
explained in these models,
exactly in the same way as 
the hierarchy of the gauge couplings of the SM
can be explained if one assumes  the existence of a unifying
gauge symmetry at $M_{\rm GUT}$ \cite{gut1}.

It has been also observed \cite{kmoz} 
that there exists a
relatively wide range of $k$'s which gives the 
top-bottom hierarchy of the right order. 
Of course, the existence of this range is partially related to the
infrared behavior of the Yukawa couplings \cite{hill1}.
 Therefore, a systematic investigation
on the nature of GYU is
indispensable
to see whether a
GYU can make experimentally distinguishable predictions on
the top and bottom masses, or whether the
top-bottom hierarchy results mainly from the infrared behavior of the
Yukawa couplings. 
With more precise measurements of the top and bottom masses,
we will be able to  conclude which case is indeed realized.

We have
performed an exhaustive analysis on this problem 
at the two-loop level \cite{kmoz},
and here we would like to
 present only a few representative results to provide an idea of our
complete analysis.
We have made the same assumptions as in chapters 3, 4, and 5, namely
that below $M_{\rm GUT}$ the evolution of couplings is 
 governed by the MSSM and that there exists a unique threshold
$M_{\rm SUSY}$ for all superpartners of the MSSM so that
below $M_{\rm SUSY}$ the SM is the correct effective 
theory, where we include only
the logarithmic and two-loop corrections
for the RG evolution of 
couplings.
We have neglected all the
threshold effects. Note that with a GYU boundary condition
alone the value of $\tan\beta$ can not be determined;
usually, it is determined in the Higgs sector, which however
strongly depends on the supersymmetry breaking terms.
In our analysis we avoid this by using the tau 
mass, along
with  $M_Z$, $\alpha_{\rm em}^{-1}(M_{Z})$ and $
\sin^{2} \theta_{\rm W}(M_{Z})$, as the input.
 Figure 3 shows
the dependence of the top mass on different values of
$\kappa _t^2$, when $\kappa _r^2 = \kappa _{\tau}^2/\kappa _b^2$ is fixed
to 2.0, and the supersymmetry breaking scale $M_{SUSY}=500$ GeV.

At this point it is also worth recalling the predictions for $m_t$
of ordinary GUTs, in particular of supersymmetric $SU(5)$ and
$SO(10)$.  The MSSM with $SU(5)$ Yukawa boundary unification allows
$m_t$ to be anywhere in the interval between 100-200 GeV \cite{barger}
for varying $\tan \beta$, which is now a free parameter.  Similarly,
the MSSM with $SO(10)$ Yukawa 
boundary conditions, {\em i.e.} $t-b-\tau$ Yukawa Unification gives
$m_t$ in the interval 160-200 GeV \cite{barbero}.
 
It is clear that the GYU scenario presented here, provides us
with predictive solutions which are consistent with all relevant
experimental data. Moreover, it is the most predictive scheme as far as
the mass of the top quark is concerned. It is worth noting
that the various GYU models can be well distinquished among themselves
when more information on the supersymmetry breaking scale is available.
A nice demonstration of this point is shown in Figure 4.

Concerning recent related studies, we would like to
emphasize that our approach of dealing with asymptotically non-free
theories \cite{kmtz} covers work done by other authors
\cite{lanross}, though the underlying 
idea might be different.  Also, interesting RGI relations
among the soft breaking parameters above the unification scale have
been found \cite{jj}.  These relations are obtained in close analogy to our
approach.

It will be very interesting to find out in the comming years, as the
experimental accuracy of $m_t$ increases, if nature is kind enough to
verify our conjectured Gauge-Yukawa Unification.

\vspace {0.8cm}

\noindent{\bf Acknowledgements}
\vspace {0.5 cm}

It is a pleasure to thank our collaborators D. Kapetanakis,
M. Olechowski, and N.D. Tracas for their contribution in the
development of the above ideas.  We would also like to thank 
M. Aschenbrenner, M. Carena, H. Leutwyler, D. Matalliotakis, 
P. Minkowski, H.P. Nilles, R. Oehme,
S. Pokorski, M. Quir\'os, G.G. Ross, C. Savoy, K. Sibold, C. Wagner, and
W. Zimmermann for useful conversations.
G.Z. would like to thank the organizers for the invitation and their
kind hospitality.

\newpage
\begin{center}
{\bf\Large Figure Captions}
\end{center}

\vspace{1cm}
\noindent
{\bf Fig.\ 1}. The comparison of the predictions for $m_t$ between
    the AFUT models and the FUT one. For
    the FUT model $\tilde{\alpha}_t=1.6$, $\tilde{\alpha}_b=1.2$, for
    AFUT3 $\tilde{\alpha}_t=0.97$, $\tilde{\alpha}_b=0.57$, and for
    AFUT4 $\tilde{\alpha}_t=1.37$, $\tilde{\alpha}_b=0.97$.

\vspace{1cm}
\noindent
{\bf Fig.\ 2}.  The values for $m_t$ predicted by the Pati-Salam model
    for different $M_{SUSY}$ scales.  Only the ones with $M_{SUSY}$
    beyond 400 GeV are realistic.

\vspace{1cm}
\noindent
{\bf Fig.\ 3}. The dependence of the top mass $m_t$ with $k^2_t$, at
    fixed $M_{SUSY}=500$ GeV.
    As we can see, after $k^2_t\sim 2.0$ the top mass goes to its infrared
    fixed point value.

\vspace{1cm}
\noindent
{\bf Fig.\ 4}.  $m_t$ predictions of $SU(5)$ FUT and AFUT3 models, for
    given $M_{SUSY}$ around 100 and 500 GeV. For
    the FUT model $\tilde{\alpha}_t=1.6$, $\tilde{\alpha}_b=1.2$, and for
    AFUT3 $\tilde{\alpha}_t=0.97$, $\tilde{\alpha}_b=0.57$.

\newpage

\begin{figure}
           \epsfxsize= 11 cm   
           \centerline{\epsffile{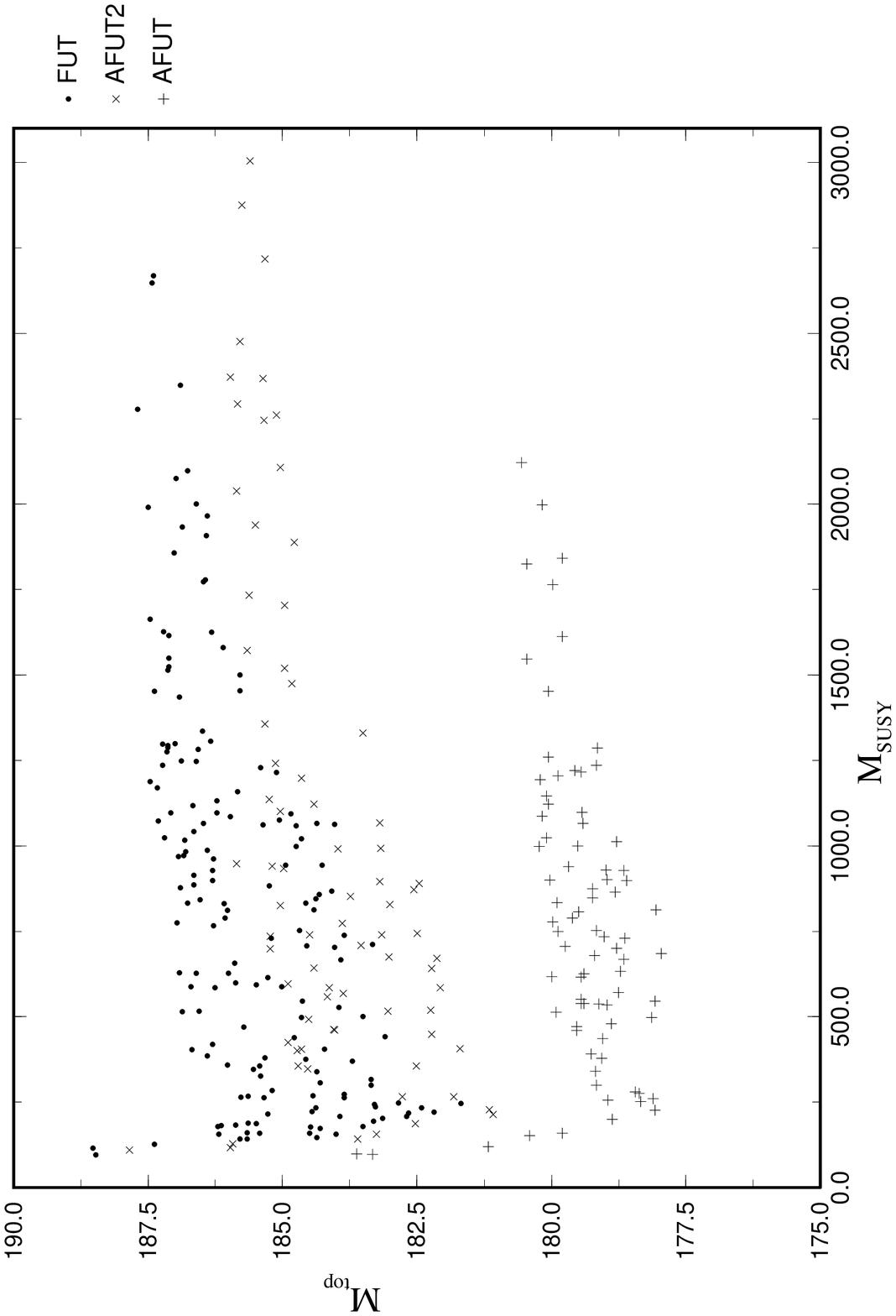}}
\label{fig:1}
\end{figure}

\newpage

\begin{figure}
           \epsfxsize= 11 cm   
           \centerline{\epsffile{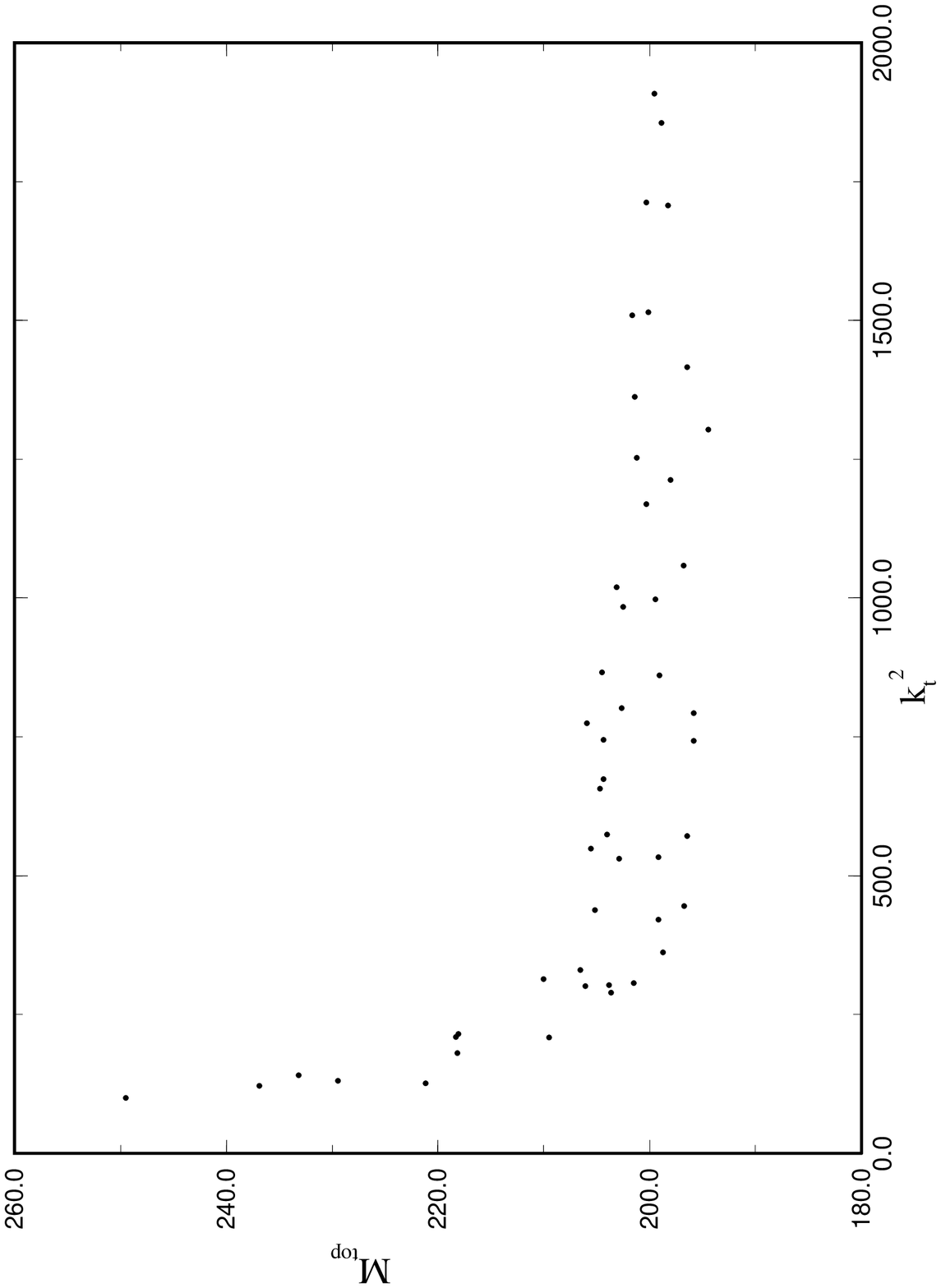}}
\label{fig:2}
        \end{figure}

\newpage

\begin{figure}
           \epsfxsize= 11 cm   
           \centerline{\epsffile{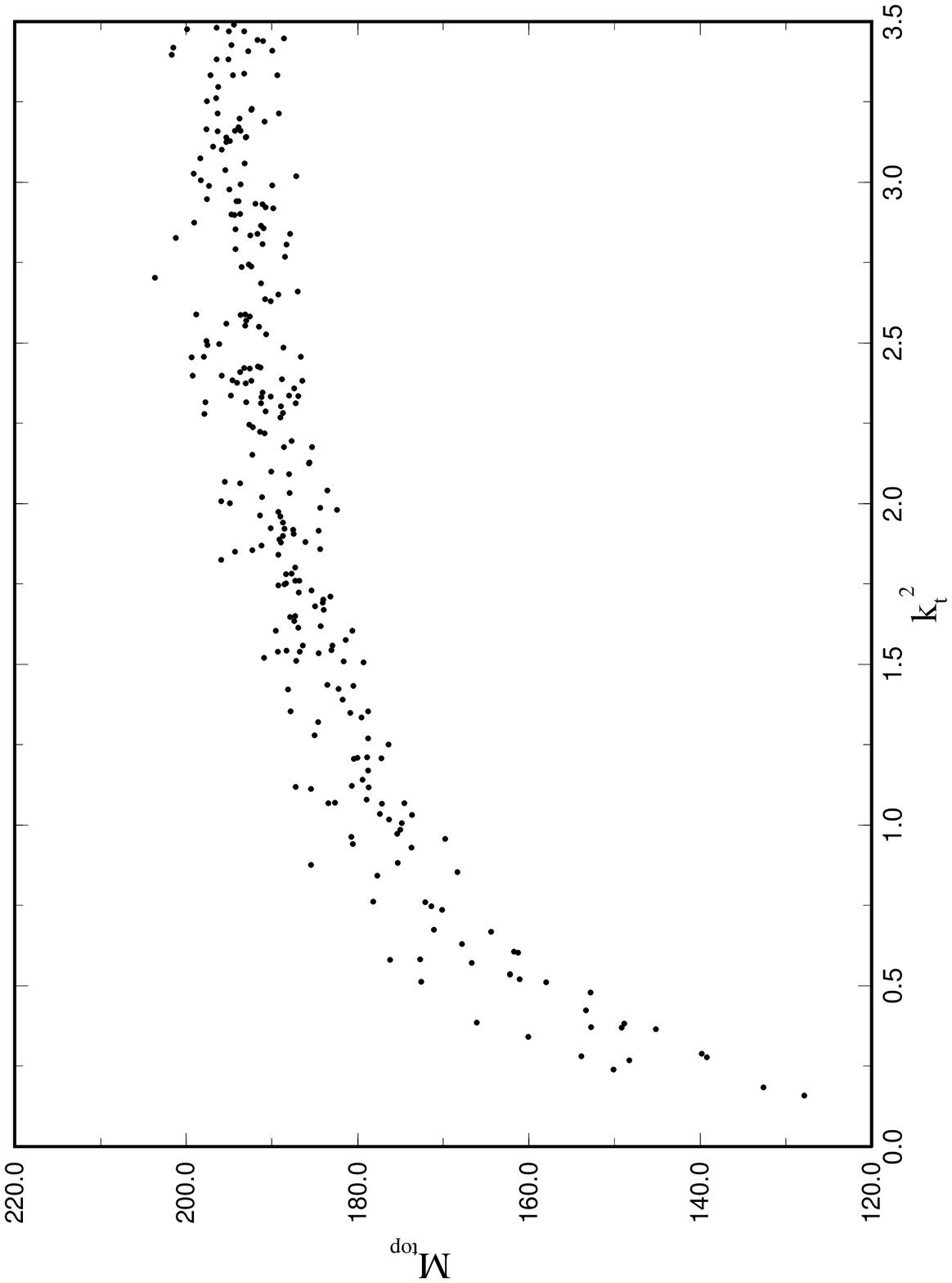}}
\label{fig:3}
        \end{figure}

\newpage

\begin{figure}
           \epsfxsize= 11 cm   
           \centerline{\epsffile{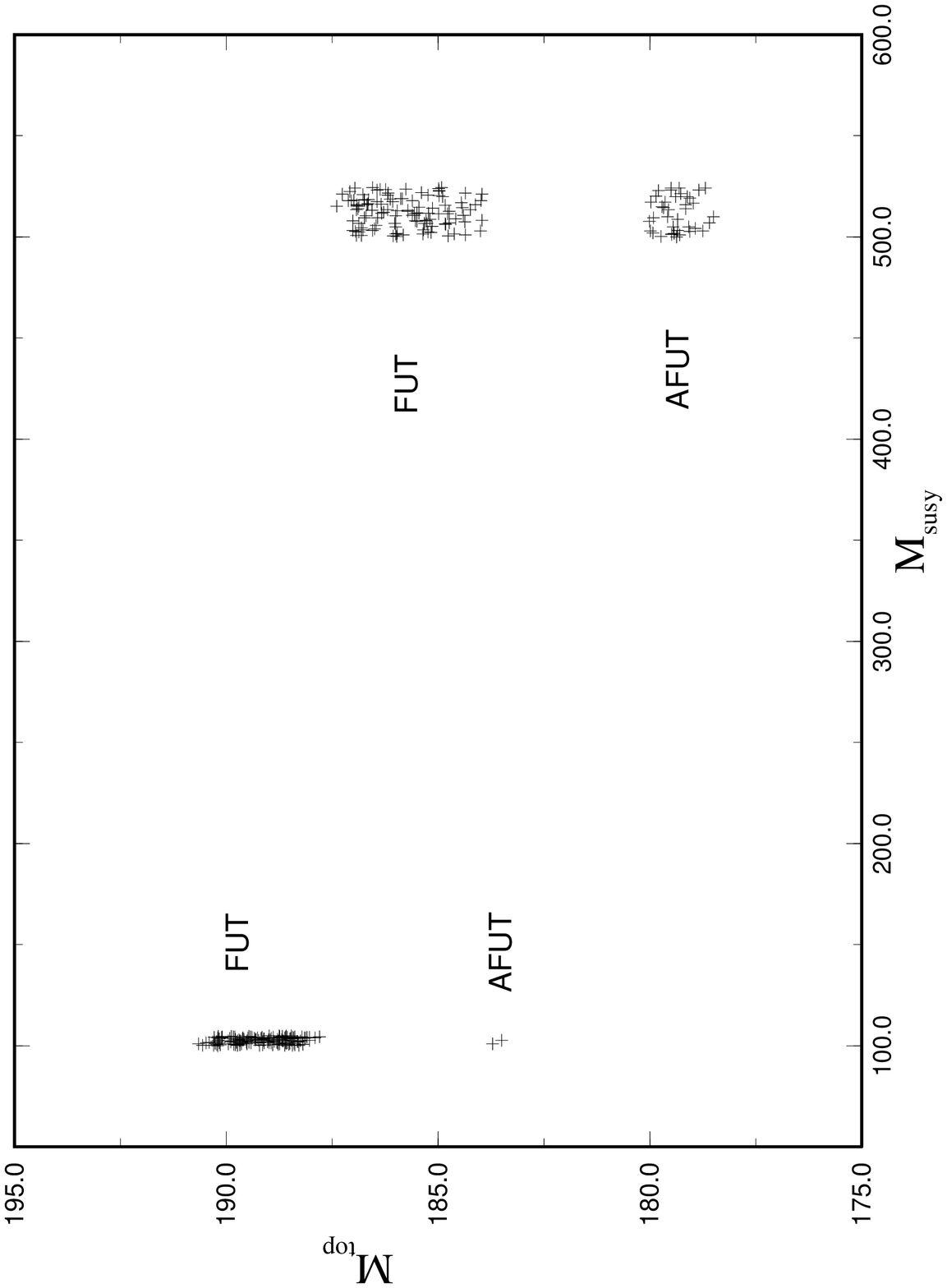}}
\label{fig:4}
        \end{figure}

\end{document}